# ENGINEERING TUNABLE STRAIN FIELDS IN SUSPENDED GRAPHENE BY MICROELECTROMECHANICAL SYSTEMS

Jens Sonntag[1,2], Matthias Goldsche[1,2], Tymofiy Khodkov[1], Gerard Verbiest[1], Sven Reichardt[3], Nils von den Driesch[2], Dan Buca[2], and Christoph Stampfer[1,2]

[1]JARA-FIT and 2nd Institute of Physics, RWTH Aachen University, GERMANY, and
[2]Peter Grünberg Institute (PGI-8/9), Forschungszentrum Jülich, GERMANY, and
[3]Department of Materials, University of Oxford, UNITED KINGDOM

## ABSTRACT

Here, we present a micro-electromechanical system (MEMS) for the investigation of the electromechanical coupling in graphene and potentially related 2D materials. Key innovations of our technique include: (1) the integration of graphene into silicon-MEMS technology; (2) full control over induced strain fields and doping levels within the graphene membrane and their characterization via spatially resolved confocal Raman spectroscopy; and (3) the ability to detect the mechanical coupling of the graphene sheet to the MEMS device with via their mechanical resonator eigenfrequencies.

## INTRODUCTION

The unique combination of electrical and mechanical properties of graphene makes it a prime candidate for flexible electronics such as bendable high-frequency resonators, sensors, filters, mixers, or amplifiers [1]. The application of strain alters the mechanical properties, e.g., the frequency of a graphene resonator, but more remarkably also the electronic properties. Strain gradients induce a pseudo-magnetic field similar to that of a real magnetic field [2]. Crucially, inhomogeneous strain fields have been identified as limiting the carrier mobility in high-quality graphene samples [3]. Control over strain fields may lead to fully strain-engineered devices such as valley filters [2] as well as piezoelectric and superconducting devices [4]. Strain is commonly induced in graphene by pulling on suspended sheets with an electrostatic gate [5] or by bending flexible substrates [6]. The obtained strain fields are thus intrinsically linked to either the electronic tuning of the charge carrier density or to the properties of the substrate. This lack of independent control over strain fields poses a great challenge for any application based on strain-engineered graphene devices. Moreover, engineering truly controllable local strain patterns in graphene has not been achieved so far.

Here we present a silicon-based platform that allows (i) to strain suspended graphene and (ii) to independently tune the charge carrier density. Our devices further allow us to locally create strain hotspots. Additional gold contacts enable us to conduct electronic transport measurements. This enables us to measure the mechanical resonance frequency of the graphene membrane as a function of strain.

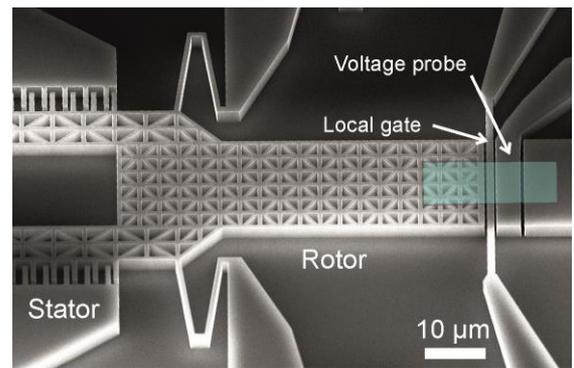

**Figure. 1: Scanning electron microscope (SEM) image of a multiterminal silicon MEMS device, capable of straining and doping graphene. The cyan area indicates the tranfer area for the graphene flake.**

## SYSTEM DESCRIPTION

Figure 1 shows the design of our silicon-based MEMS device for engineering strain fields in graphene, which is based on the design presented in Refs. [7-9]. The actuator consists of two interdigitated combs. One of these combs, together with the entire rotor structure is suspended above the substrate via springs. By fabricating the fingers asymmetrically and applying a voltage $V_{CD}$ between rotor and stator, a force

$$F = \eta \cdot V_{CD}^2 \qquad (1)$$

is applied to the comb-drive (CD), which results in a displacement $d = k \cdot F$. Here, $\eta$ is a constant related to the capacitance of the combs and $k$ is the spring constant of the combined graphene-clamping-silicon geometry. The actuators presented here are improved compared to our previous works [7,8] in two respects: (i) we include additional gold-coated contact areas for multi-terminal transport measurements and (ii) we optimized the clamping geometry (see Figure 2) for creating local strain hot spots and strain gradients, which is imperative for the observation of pseudomagnetic fields and valley-effects [2]. To conduct transport measurements, we also fabricate a local gate, over which the graphene flake is suspended. This allows a full control over both strain and doping within the graphene flake. Furthermore, it allows the excitation of motion into the graphene flake and enables the operation as resonator.

The fabrication of these devices follows the procedure previously reported in Refs. [7-9]. We start with a silicon-on-insulator (SOI) substrate with a 2 µm thick, chemical vapor deposition-grown, highly doped silicon layer. Standard electron beam lithography is used to manufacture either a chromium or an aluminum hard mask. We then use reactive ion etching with $C_4F_8$ and $SF_6$, to pattern the actuators. If a local gate is employed, the etching process is stopped after a depth of 300 nm. A consequent lithography step creates a hard mask for the local gate, after which the remaining silicon layer is removed. The integration of graphene flakes into the silicon-based technology is done by exploiting the PMMA used to transfer the graphene onto the device. By means of electron beam lithography, the PMMA is locally cross-linked, creating a fixed connection between the silicon and graphene. After dissolving the remaining PMMA, we release the comb-drive in hydrofluoric acid. A critical point-drying step prevents the collapse of the suspended structure.

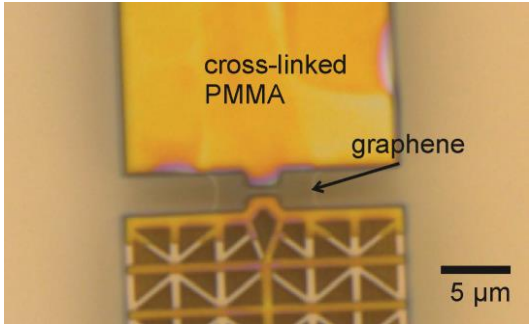

**Figure 2:** Optical image of a transferred graphene flake on a released comb-drive used to locally induce high amounts of strain and strain gradients.

Figure 2 shows our approach to locally achieving high amounts of strain by including an improved 'double nose' clamping geometry in our actuator design. This symmetric approach further localizes the strain. This allows an increase of the achievable maximum strain, as in this design, the strain at the edges of the graphene flake is comparatively low and the edges are the areas most sensitive to rupture [7].

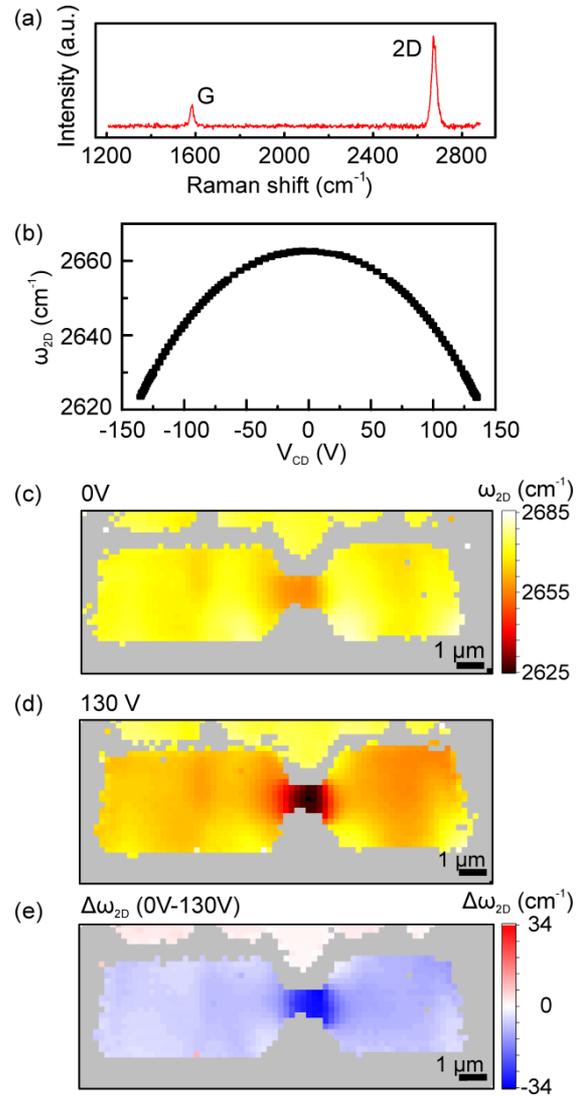

**Figure 3:** (a) Raman spectrum of graphene; (b) parabolic dependence of the 2D peak position on $V_{CD}$ as predicted for strain; (c) & (d) spatially resolved Raman maps of the 2D peak position for actuator voltages of 0 V and 130 V, respectively; (e) change in $\omega_{2D}$ from $V_{CD} = 0$ V to $V_{CD} = 130$ V, showing the tunable, localized strain field.

# EXPERIMENTAL RESULTS
## Inducing strain and strain hotspots

We characterize the tunable strain fields by spatially resolved confocal Raman spectroscopy [10]. Figure 3a shows the Raman spectrum of graphene, which consists of the so-called Raman G and 2D peaks. We extract the peak-positions $\omega_G$ and $\omega_{2D}$ by fitting two Lorentzian peaks to each taken spectrum. $\omega_G$ and $\omega_{2D}$ are highly sensitive to strain und shift by 37.7 cm$^{-1}$/% and 83 cm$^{-1}$/% per percent of applied uniaxial strain, respectively [6]. As the induced strain is proportional to the applied force (Eq. 1), we expect a quadratic dependence of the 2D peak position $\omega_{2D}$ as a function of applied voltage $V_{CD}$. This quadratic dependence depicted in Figure 3b proves that the peak shift is purely induced by straining graphene and not by doping, as this would give rise to a non-symmetric shift. The combined system of graphene and MEMS allows a highly reproducible and controllable operation, as there is no hysteresis visible in the 2D peak positions, as shown in Figure 3b. Figure 3c-e shows spatially resolved Raman maps of a graphene flake integrated in our MEMS device. By applying a voltage $V_{CD}$ we are able to create a strain hotspot and corresponding strain gradients within the graphene flake. The resulting peak position gradient is on the order of $\partial \omega_{2D}/\partial x = 35 \text{ cm}^{-1}/\mu\text{m}$, which corresponds to a strain gradient of $\partial \epsilon/\partial x \approx 0.4 \%/\mu\text{m}$. Assuming that the strain is applied in the armchair direction, this gradient would induce a pseudomagnetic field of $B_{ps} = \hbar\beta/2ae \cdot \partial \epsilon/\partial x \approx 31$ mT, where $\beta = 3.37$ and $a = 1.42$ Å is the nearest-neighbor C-atom distance in graphene [2,7].

## The role of slack and the stability of the clamping

Due to the nature of the process of transferring the graphene flake, some samples show slack, folds, and/or wrinkles. Consequently, there is a certain amount of displacement where the comb-drive moves in order to remove slack or to straighten wrinkles, while the graphene lattice is effectively not strained. As Raman spectroscopy probes lattices vibrations, this straightening is not visible in Raman spectroscopy. An example of such an occurrence of slack in a bilayer graphene sample (without 'noses') is depicted in Figure 4a. Because bilayer graphene exhibits a more complex 2D peak, we show the position $\omega_G$ of the G peak, which, while slightly less sensitive to strain, carries a similar amount of information. The slack is clearly visible, as there is a certain voltage range and thus a certain displacement where $\omega_G$ does not change, i.e., where the graphene lattice is not strained. Only after all the slack is overcome does the lattice deform, which then alters the Raman signal and restores the parabolic dependence of $\omega_G$ as a function of $V_{CD}$.

While this 'intrinsic' slack is rare in our samples, we also observed that in some samples, where slack arises from the measurement itself. Such a behavior of a sample with a 'nosed'-clamping geometry is shown in Figure 4b. Here we depict a series of measurements in which we sweep $V_{CD}$ from $-V_{max}$ to $V_{max}$ and incrementally increase $V_{max}$ from 25 V to 40 V (see labels in Figure 4b). While for small $V_{max}$ there is no apparent slack, the repetition of the measurement and the increase of the amount of strain leads to the formation of a clearly observable slack, indicated by the deviation of $\omega_{2D}$ from the expected parabolic behavior (shown as a red line in Figure 4b).

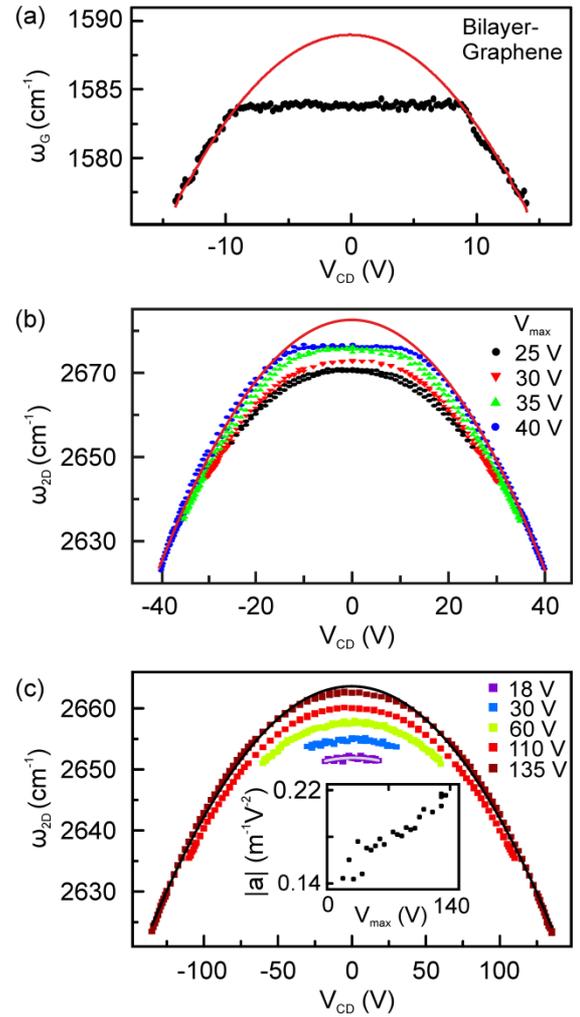

**Figure 4: (a)** G peak position of a bilayer graphene sample as a function of $V_{CD}$ showing an intrinsic amount of slack. **(b)** The 2D peak position of a monolayer graphene sample as a function of $V_{CD}$ shows an increasing amount of slack, when $V_{max}$ is increased. **(c)** The parabolic dependence of $\omega_{2D}$ of a different monolayer sample shows an higher amount of tunability with increasing $V_{max}$. The inset shows the increase in curvature of the fitted parabola.

One possible origin for this slack could be a limited amount of plastic deformation of the crosslinked PMMA acting as clamp. We can estimate the amount of plastic deformation by the discrepancy between the fitted parabola and the data at $V_{CD} = 0$ V. Under the assumption that the spring constant of the coupled silicon-graphene system does not change significantly, we can calculate the displacement of the comb-drive from the shift in $\omega_{2D}$ by calculating the strain $\epsilon$ via the strain-induced shift of 83 cm$^{-1}$/% [6]. Subsequently, we calculate the displacement $d = \epsilon \cdot l$ by considering the distance $l = 750$ nm between the two clamping sides at the position of the nose. We find a deviation of $\omega_{2D}$ from the parabolic behavior at $V_{CD} = 0$ V of approximately 8 cm$^{-1}$. This translates to a plastic deformation of the PMMA clamping of below 1 nm. Such a deformation is reasonable to assume, considering the high amount of strain and the enormous stiffness of the graphene of 1 TPa [11]. A secondary effect of the plastic deformation is the relaxation of the pre-strain within the graphene lattice without any applied voltage ($V_{CD} = 0$ V). This pre-strain (see also Figure 3c) typically exists within our sample because of the contraction of the PMMA during cross-linking. We only observe the developing of slack in samples with localized strain hotspots and only at high strain values.

An additional effect can be observed when $V_{max}$ is increased: the curvature of the parabola increases. This is most visible in Figure 4c, which shows $\omega_{2D}$ as a function of $V_{CD}$ in a different device with a 'nosed' clamping geometry. Note that this device does not show the development of slack. There is however a clear increase in tunability, which we quantify by fitting $\omega_{2D}$ with a parabola $\omega_{2D} = a \cdot V_{CD}^2 + \omega_0$ and by extracting the curvature $a$. As seen in the inset in Figure 4c, there is an increase in curvature $a$ of nearly 60%. We attribute this behavior to an increased $\eta$, i.e., an increase of the capacitive force. Due to the fact, that the graphene is the strongest spring in our system, the cross-linking and the resulting pre-strain leads to a displacement of the comb-drive in the direction opposite to its intended move direction. This leads to a reduction of the finger asymmetry and net force in the initial state. High enough strain relaxes part of this cross-linking pre-strain and might also flatten out wrinkles within the graphene, which increases the asymmetry and thus the force.

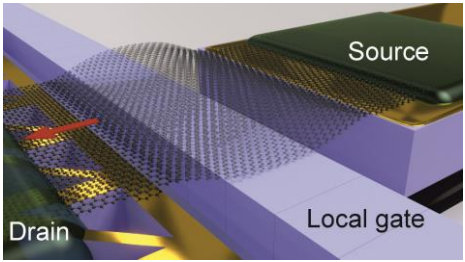

**Figure 5: Schematic illustration of a graphene resonator integrated into a MEMS device.**

**Graphene resonator coupled to silicon-MEMS**

The developed technique is also suitable for on-chip applications, as we can also detect the strain via the resonance frequency of the suspended graphene membrane, when operated as mechanical resonator (see Figure 5). By using the local gate, we excite a motion of the graphene flake and use a current flowing through the graphene to detect the resonance frequency in an amplitude-modulated down-mixing scheme [5]. In this scheme, we apply a DC voltage $V_{LG}$ as well as an AC voltage at frequency $f$ to the local gate. In addition, a small modulation voltage $V_{SD} = 10$ mV at frequency $f \pm \Delta f$ is applied across the graphene membrane. Due to change of the conductance $G$ of the graphene flake with distance from the gate and the gate tunability $\partial G/\partial V_{LG}$, the motion of the graphene membrane is detected as a down-mixing current $I_{DM}$ at frequency $\Delta f$. The current $I_{DM}$ is amplified with an IV-converter and measured with a lock-in amplifier. When the frequency $f$ is tuned to the mechanical resonance of the graphene sheet, $I_{DM}$ shows a distinct resonance, which we then fit to extract the resonance frequency.

Figure 6 shows the extracted resonance frequency as a function of $V_{CD}$. We again observe a parabolic tuning of the resonance frequency in agreement with the applied electrostatic force. Due to capacitive softening, i.e., the electrostatic force between the graphene membrane and the local gate, we can also tune the resonance frequency via the applied DC gate voltage $V_{LG}$ (shown as a dashed line in Figure 6).

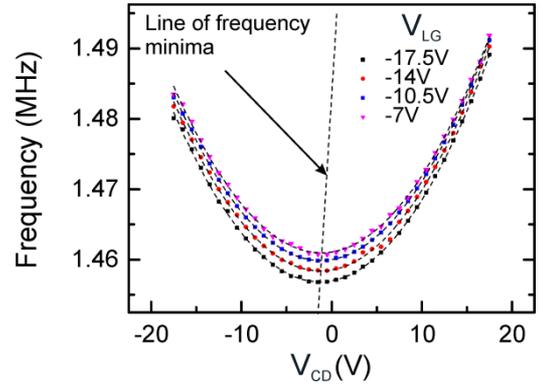

**Figure 6: Tunable resonance frequency of the combined graphene-MEMS device as a function of applied actuator potential. The dotted line indicates a tuning of the frequency with applied $V_{LG}$.**

## CONCLUSIONS

We designed and fabricated silicon-based comb-drives for studying the electromechanical coupling in graphene. We are able to successfully clamp graphene to the silicon device and strain the graphene membrane in a controllable way. By employing spatially resolved Raman spectroscopy we verified our clamping concept to create local strain hotspots. The implementation of a local gate and gold contacts we are also able to probe electrical transport through the graphene. Here, we employ this to electrically probe the mechanical resonance frequency of the graphene membrane. We expect that the presented platform will prove valuable in a wide range of applications and for further studies of the coupling of electronic and mechanical degrees of freedom, not only in graphene, but also other two-dimensional materials.

## ACKNOWLEDGEMENTS

Support by the ERC (GA-Nr. 280140), the Helmholtz Nano Facility [12] and the Deutsche Forschungs-gemeinschaft (STA 1146/12-1) are gratefully acknowledged. This project has received funding from the European Unions Horizon 2020 research and innovation programme under grant agreement No 785219.

## CONTACT

*J. Sonntag, sonntag@physik.rwth-aachen.de